# Space-Time Aspects of Quasiparticle Propagation

Richard Levien[*]

Chetan Nayak[†]

*Department of Physics*
*Joseph Henry Laboratories*
*Princeton University*
*Princeton, N.J. 08544*

Frank Wilczek[‡]

*School of Natural Sciences*
*Institute for Advanced Study*
*Olden Lane*
*Princeton, N.J. 08540*

---

[*] levien@pupgg.princeton.edu
[†] Research supported in part by a Fannie and John Hertz Foundation fellowship. nayak@puhep1.princeton.edu
[‡] Research supported in part by DOE grant DE-FG02-90ER40542. WILCZEK@IASSNS.BITNET

# ABSTRACT

Highly correlated states of electrons are thought to produce quasiparticles with very unusual properties. Here we consider how these properties are manifested in space-time propagation. Specifically, we discuss how spin-charge separation is realized in a simple double-layer geometry, how mass renormalization affects time-of-flight in compressible Hall states, and how quantum drifts can reveal the effective charge, mass, and quantum statistics in incompressible Hall states. We also discuss the possibility of observing the effect of fractional statistics directly in scattering. Finally we propose that, as a result of incompressibility and the fundamental charge-flux relation, charged probes induce macroscopic, measurable rotation of Hall fluids.

# 1. Introduction

Over the past few years, qualitatively new behaviors of interacting electrons have been observed in deep quantum regimes involving some combination of low temperature, low dimensionality, and large magnetic fields. The theoretical description of these states predicts that their low-energy excitations can be described as quasiparticles, which as their name implies have particle-like properties – but with several unusual twists including in various cases fractional charge, fractional statistics, singular mass renormalization and spin-charge separation. Although the theory as a whole is compelling, and consequences of it have been tested both by experiments and by extensive numerical work, for the most part these dramatic predictions for exotic properties of quasiparticles have been exhibited only rather indirectly. As experimental technique and sample purity improve, one might anticipate that conceptually ideal experiments that directly probe ballistic quasiparticle propagation will become feasible. In this note, we will explore a few especially interesting issues that arise in interpreting such propagation. We also suggest a possible macroscopic manifestation of the charge-flux relation for quasiparticles, involving the rotation of the entire Hall fluid when an external charge is brought near.

# 2. Spin-Charge Separation

The separation of spin and charge quantum numbers is a phenomenon of fundamental interest. It is firmly predicted to occur in realistic 1+1-dimensional models involving solitons (*e. g.* polyacetylene) or non-trivial long wavelength dynamics (*e. g.* 1d metals). It is also known to occur in a wide variety of field-theory models, including some in higher dimensions, and has been vigorously discussed as a possibility for the normal state of copper oxide superconductors [1]. Here we propose a simple, direct, and seemingly accessible space-time manifestation of quantum number separation in layered Hall systems.

Consider a system of two nearby layers of electron fluid, subject to a strong perpendicular magnetic field. One knows that in suitable circumstances the ground state of the electrons is an incompressible correlated state. The low-energy excitations of a bounded sample are then described as surface waves of the (two-layered) electron droplet. We will consider the case of a simple edge, such that there is just one mode in each separate layer. The dynamics of these waves, at long wavelengths, is governed by the 1+1-dimensional field theory with a Lagrangian density of the form [2]

$$L \ = \ K^{ij} \partial_t \phi^i \partial_x \phi^j \ + \ V^{ij} \partial_x \phi^j \partial_x \phi^j \tag{2.1}$$

In this expression $i, j$ are labels for the layer (or quasi-spin) and $K, V$ are symmetric numerical matrices. (In principle they could depend on $x$ or $t$, but this complication is inessential for us.) This is the universal Hamiltonian symmetric between the layers, containing the leading terms at long wavelength. The $\phi^i$ are invariant under addition of a constant, which implies number conservation for the charges $\int \partial_x \phi^i$. The form of $K$ is determined by the bulk theory. The $V$ matrix is not universal. One source for it, presumably mainly supplying the diagonal terms, is $E \times B$ drift (the off-diagonal terms in $K$ reflect the fact that this interaction serves to induce the velocities characteristic of drift in the *effective* magnetic fields, which include contributions from correlation phases). $V$ is also affected, presumably mainly in its off-diagonal pieces, by ordinary Coulomb repulsion.

In the simple case that the layers are symmetric, so $K_{11} = K_{22}$ and $V_{11} = V_{22}$, the propagating modes with definite velocities are made up from $\phi^1 \pm \phi^2$. The upper sign represents a *holon* – charge without spin – while the lower represents a *spinon* – spin without net charge. They propagate at velocities

$$\begin{aligned} v_{\text{holon}} &= \frac{V_{11} + V_{12}}{K_{11} + K_{12}} \\ v_{\text{spinon}} &= \frac{V_{11} - V_{12}}{K_{11} - K_{12}} \ . \end{aligned} \tag{2.2}$$

Now consider the effect of injecting an electron at the edge of one layer, as exhibited in Figure 1. It will decompose into a spinon and holon, which propagate at different speeds. Downstream of the injection point, then, one will first see a pulse of voltage difference between the two layers, with no net charge, and then a pulse of charge, with no voltage difference. (This is for $V_{12}$ small. For $V_{12}$ large, the order will be reversed.) This is spin-charge separation in a very tangible form.

A nice example for this framework is provided by the Halperin-Laughlin (m,m,n) states [3]. They have $K_{11}/K_{12} = m/n$. Thus in the limit where $V$ is approximately diagonal (that is, for sharp field gradients near the edge, or weak coupling between the layers) the ratio of holon to spinon velocities is expressed directly in terms of the fundamental integers characterizing the state, *i.e.*

$$\frac{v_{\text{holon}}}{v_{\text{spinon}}} \rightarrow \frac{m+n}{m-n} \ . \tag{2.3}$$

Under certain circumstances one can induce phase transitions between states at different values of $m, n$ but with the same filling fraction $2/(m+n)$; our considerations imply that there are qualitative changes in the velocity spectra through the transition.

## 3. Mass Renormalization and Time-of-Flight

Recently Halperin, Lee, and Read [4] proposed a striking description of the compressible quantum Hall state near filling fraction $\nu = 1/2$, which has gained some experimental support. Their theory is based on the idea, which sounds startling on first hearing, of approximating the electrons by fermions in *zero* magnetic field. The rationale for this idea is that correlations between the electrons are such as to cancel the effect of the imposed magnetic field, so that the dressed quasiparticles have the quantum numbers of electrons but travel in straight lines. The special feature of $\nu = 1/2$ is that attaching notional two-unit flux tubes to electrons formally does not alter their properties, yet it involves a vector potential which cancels off the growing part of the real imposed vector potential. Thus one might hope to treat the difference as a perturbation, especially since it is at least quasi-local and the Fermi liquid one is perturbing away from, though gapless, is quite robust.

The perturbative treatment of the statistical gauge field is intrinsically limited, however, because the coupling is neither small nor truly short-range. Nor, since it is essentially a magnetic coupling, is it effectively screened. Two of us have presented a renormalization group analysis of this problem, with the following main result (for alternative discussions, see [6]. In [5], we described our results in equivalent but more technical language). The dangerous interaction is sensitive to long-wavelength density fluctuations. In considering these fluctuations, it is important to include the effect of the Coulomb repulsion (which suppresses them); for the sake of generality, one considers a $1/|k|^x$ interaction, where $x = 1$ is the Coulomb case. Among the couplings in a non-relativistic Lagrangian is the mass term $\frac{1}{2m^*}(\nabla\psi)^2$, and it turns out that the crucial renormalization for the gauge theory is the renormalization of $m^*$ as one one scales toward energies and momenta on the Fermi surface. Indeed since the fundamental coupling is magnetic, an increase in $m^*$ tends to suppress it. For general $x$ we find a fixed point for the effective coupling at $\alpha^* = (1-x)/4$; for the Coulomb case there is a logarithmic approach to zero coupling. For the effective mass in this case we have

$$m^*(\omega) \to \text{const.} \ln\frac{1}{|\omega|} \tag{3.1}$$

as one approaches the nominal Fermi surface. Notice that this mass diverges at the Fermi surface, so that strictly speaking the Fermi liquid theory does not apply, though many of its qualitative features survive in a modified analysis. Clearly the predicted behavior (3.1) is of fundamental interest, as is the question whether it can be seen directly in space-time.

Actually to do this would seem to require only that the beautiful measurements of Goldman, Su, and Jain [7] be extended. These authors tested some fundamental properties of the quasiparticles by working in magnetic fields $B_{1/2} + \delta B$ slightly different from those which give exactly $\nu = 1/2$. The quasiparticles are then supposed to see the effective field $\delta B$, which causes them to move in large cyclotron orbits. In fact since the cyclotron radius is $r = pc/(e\delta B)$ and the momentum is cut off at the Fermi surface there is predicted to be a minimum radius, a prediction which was verified experimentally using a two-slit arrangement. Now what is required to test (3.1) is clearly that the *time of flight* should be measured as a function of the radius; from it one readily obtains the velocity and, knowing the momentum, the effective mass.

## 4. Quantum Drifts: Effective Field, Charge, Mass, and Statistics

We would like now to discuss how the properties of quantized Hall effect quasiparticles are reflected in their ballistic propagation in slowly varying external fields. In the incompressible Hall states, as opposed to the $\nu = 1/2$ state just discussed, the effective field seen by the quasiparticles is not zero (though in general it is modified from the imposed field by definite, substantial multiplicative factors, for example a factor $1/3$ in the classic $\nu = 1/3$ Laughlin state.) The calculation of the motion of charged particles in a strong, nearly homogeneous magnetic field and additional small perturbations is a classic chapter of plasma physics [8]. The leading idea is that to a first approximation in strong magnetic field the motion is simply circular cyclotron motion with the characteristic frequency $\omega_c = \frac{qB}{mc}$, and radius

$$r_{\mathrm{cl}} = \frac{pc}{qB} . \tag{4.1}$$

When there are additional weak forces that vary slowly in time and space (on the scales defined by $\omega_c$, $r_{\mathrm{cl}}$) then in addition to this fast motion one finds slow drifts. The drifts may be calculated by averaging the dynamical equations over the fast motion, and finding the residual motion. Now when quantum mechanics is taken into account one finds that there is another natural length scale, the magnetic

length

$$l_{\rm B} \equiv \sqrt{\frac{\hbar c}{qB}} \ . \tag{4.2}$$

When this magnetic length becomes comparable to the length scale over which the weak fields vary, clearly the classical averaging procedure is inadequate and one must use a different method to find the drift motions. These quantum drifts will allow one to measure the charge and mass of the quasiparticles separately. We will present results for the quantum drifts of a particle with charge $q$ and mass $m$ subject to magnetic field $B$. We note that in general the magnetic field $B$ seen by a quasiparticle is *not* the same as the external magnetic field, and of course the charge $q$ is some rational multiple of $e$: the microscopic theory suggests definite values for $q$ and $B$ at each filling fraction.

Formalisms have been elaborated to deal with motion in this deep quantum regime, but they are quite complex and do not easily lead to explicit results beyond the leading order. The fundamental difficulty, which makes it plausible that this complexity is intrinsic, is the vast degeneracy present in the unperturbed problem. One can readily obtain results of interest for our purposes by specializing to the case of circular symmetry in two spatial dimensions. This specialization vastly simplifies the problem, because there is at most one state of a given angular momentum in each Landau level. Thus in calculating corrections due to circularly symmetric perturbations one is doing non-degenerate perturbation theory.

A detailed account of the calculations will be presented elsewhere; here we record a few key results.

First consider the effect of a constant radial electric field, yielding the potential $V = qEr$. One then finds for the radius, current, and velocity in the $l$th partial

wave:

$$\langle r \rangle = r_0 \left[ 1 + \frac{3}{4} \frac{l_B^2}{r_0^2} + \frac{mE}{qB^2 r_0} + \cdots \right]$$
$$J_\theta = \frac{q}{2\pi r_0} \left[ -\frac{E}{B} + \frac{3}{4} \frac{E}{B} \frac{l_B^2}{r_0^2} + \cdots \right] \quad (4.3)$$
$$v_\theta = -\frac{E}{B} - \frac{mE^2}{qB^3 r_0} + \cdots ,$$

where $r_0 \equiv \sqrt{2l}\, l_B$ is the radius at which the unperturbed lowest Landau level wavefunction is a maximum. $J_\theta$ is the integrated current through a line $\theta =$ constant, and $v_\theta$ is simply $\frac{1}{q} J_\theta \times 2\pi \langle r \rangle$. There is actually a double expansion in equation (4.3): firstly the usual perturbation expansion in powers of $V$; and secondly an expansion in $\frac{l_B}{r_0}$, which tells us how well localized the unperturbed wavefunction (which has radial width $\sim l_B$) is. Note that the leading term in $v_\theta$ is the usual $E \times B$ drift, whose form is basically fixed by Galilean invariance. The second term is in fact a classical correction (there are no $\hbar$'s!), the so-called "polarization drift", which allows one to measure the ratio $\frac{q}{m}$. Quantum corrections appear in the directly measurable quantities $\langle r \rangle$ and $J_\theta$, though to this order they cancel in the product $v_\theta$.

We also find a drift induced by electric field inhomogeneities, $v_\theta = -\frac{7}{4} \frac{E' l_B}{B} \frac{l_B}{r_0}$, where $E' = \frac{dE}{dr}\big|_{r=r_0}$. This drift, which is intrinsically quantum mechanical (i.e. it vanishes in the $\hbar \to 0$ limit), allows one to measure directly the quasiparticle charge! This is impossible classically since the charge and mass enter the equations of motion only in the combination $\frac{q}{m}$. By measuring the drift due to an electric field gradient and combining it with a measurement of the constant electric field polarization drift above, one can determine both the charge and mass of quasiparticles.

Although we shall not attempt to design experiments in any realistic detail, it seems appropriate briefly to describe extremely simple and powerful methods to produce, guide, and detect the quasiparticles. An idealized *production* process, shown in Figure 3a, follows from the fact that the quasiparticles embody the lowest energy deviations from charge homogeneity. Thus if a sharply localized external charge is brought near the incompressible Hall fluid, it will be energetically favorable to produce quasiparticles of the appropriate charge, to take advantage of the imposed field (see below). If the external charge is then moved away, the produced quasiparticles will be free to move. An idealized *guidance* process, shown in Figure 3b, follows from the fact that the quasiparticles carry charge and are subject to a large (effective) magnetic field. To a first approximation, their motion is simply the classic $E \times B$ drift. Thus the quasiparticles will tend to move along electric equipotentials. One can create approximate equipotentials, and thus preferred paths for quasiparticle motion, by bringing a conductor of the desired shape nearby, in the presence of sources that would otherwise create a smoothly varying $E$ field. By providing entry and exit arms of known orientation emerging from a scattering region, as depicted in Figure 3b, one could measure scattering probabilities as a function of angle, and thus (for example) check for the effects of the predicted exotic quantum statistics. Perhaps the most notable such effect is the "anyon swerve": the cross-section is markedly different at scattering angles $\phi$ and $-\phi$. This effect, and others of interest, are beautifully illustrated in the explicit Mott scattering cross-section (including exchange) calculated in [9]. With production occurring at a known location at a known time, and subsequent space-time propagation at least roughly under control, the problem of *detecting* the arrival of electric charge, or voltage, at a known place and time perhaps becomes manageable.

A further note is in order regarding the production process for quasiparticles, which as we have said above could involve bringing an external charge close to the Hall surface. As we create one or more quasiparticles at the origin and fix them there (by not removing the external charge), other quasiparticles (or ordinary electrons) nearby will experience two kinds of azimuthal drift: firstly due to any uncanceled Coulomb charge, and secondly due to the fictitious flux tubes attached to the newly created quasiparticles. By comparing the orbital period of say quasiparticles and ordinary electrons, it is in principle possible to disentangle these two effects, so that the charge and statistics of the quasiparticles may be separately measured.

The effect of bringing up a localized charge and inducing a fictitious flux tube has a much wider implication than for the motion of quasiparticles, however. Indeed, it induces a macroscopic angular momentum for the entire droplet. Given the connection between charge and fictitious flux, one may argue this at an intuitive level from Faraday's law, as in [10]. It is instructive, however, to compare a more microscopic argument.

The Laughlin-Vandermonde wavefunctions for incompressible droplets of Hall fluid at filling fractions $\nu = 1, 1/3$ are of the form

$$\begin{aligned} \psi_1 &= \det\{z_r^{c-1}\} e^{-\frac{1}{4}\sum |z_k|^2} \\ &= \prod_{k<l: k,l=1}^{N} (z_k - z_l) e^{-\frac{1}{4}\sum |z_k|^2} \end{aligned} \quad (4.4)$$

$$\begin{aligned} \psi_3 &= \left(\det\{z_r^{c-1}\}\right)^3 e^{-\frac{1}{4}\sum |z_k|^2} \\ &= \prod_{k<l: k,l=1}^{N} (z_k - z_l)^3 e^{-\frac{1}{4}\sum |z_k|^2} \ . \end{aligned} \quad (4.5)$$

In these expressions, $r$ is the row index and $c$ is the column index for an $N \times N$ matrix. These wavefunctions define droplets of density $2\pi$ respectively $2\pi/3$ centered around the origin, where the unit of length is the magnetic length $l_e$ for an electron, defined through $l_e^2 = \hbar c/eB_{\text{ext}}$ where $B_{\text{ext}}$ is the applied magnetic field. These constructions involve levels whose wavefunctions are ordered in powers of $z$, that describe concentric rings around the origin. To obtain the wave-function of an annulus, as would be induced by bringing an external charge $Q$ nearby, one simply starts with an appropriate non-zero power of $z$. This corresponds to leaving the inner concentric rings empty. Thus one applies an annulizing factor

$$\text{Annulizing factor} \ = \ \prod_k z_k^{Q/\nu} \ . \quad (4.6)$$

This factor, which is just what one would guess for the accumulation of many quasiholes, literally carves out a geometric hole in the fluid. It involves a large change in the angular momentum, since powers of $z$ but not of its complex conjugate $z^*$ occur.

The general relationship between the change in the angular momentum and the applied charge is

$$\Delta L = \hbar N \frac{Q}{e\nu} \qquad (4.7)$$

where $N$ is the number of particles in the droplet, and $Q$ the charge applied. The rigidity of the wave function, or equivalently the physical incompressibility of the droplet, plays a fundamental role in insuring the validity of (4.7). Since both sides of the equality (4.7) are macroscopic quantities, it would seem to be possible to test (4.7) relatively simply, using suspended or levitated samples.

In any case, we have shown how a variety of predicted exotic behaviors of quasiparticles in the quantum Hall complex of states manifest themselves directly in elementary space-time processes.

Acknowledgements: R.L. would like to thank Salim Nasser for helpful discussions.

Figure Captions.

Figure 1. An electron injected at the left on the upper edge will decay into a neutral pseudo-spin excitation (middle) and a spinless charged excitation (right).

Figure 2. A one-loop diagram which renormalizes the fermion mass.

Figure 3. The creation (a) and guidance (b) of quasiparticles. In (a) a charged needle is brought close to a Hall surface to excite a quasiparticle. In (b) the Hall surface is parallel to the plane of the page, and there is a magnetic field coming out of the page.

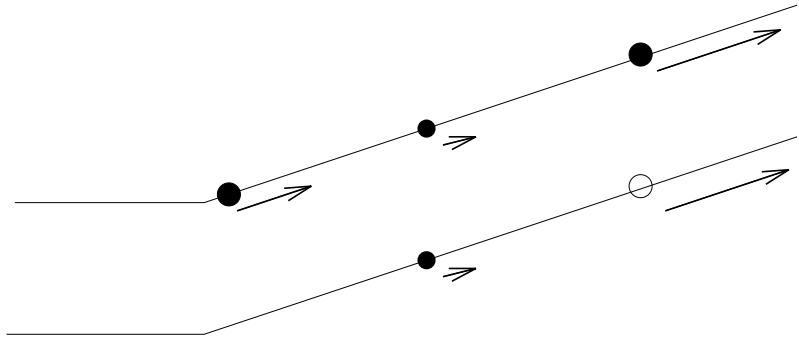

Figure 1.

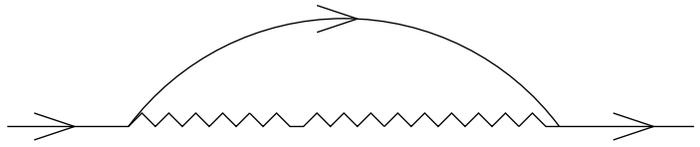

Figure 2.

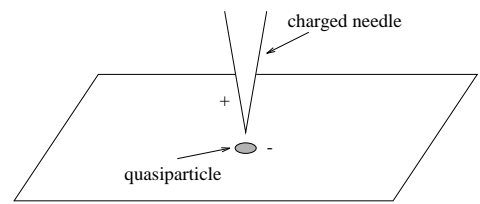

(a)

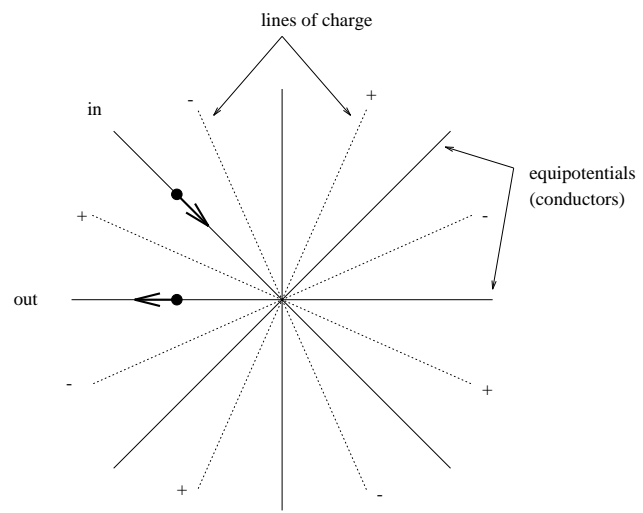

(b)

Figure 3.